# Peregrine Rogue Wave dynamics in the continuous nonlinear Schrödinger system with parity-time symmetric Kerr nonlinearity

Samit Kumar Gupta and Amarendra K. Sarma*
*aksarma@iitg.ernet.in

Department of Physics, Indian Institute of Technology Guwahati, Guwahati-781 039, Assam, India.

**Abstract -** In this work, we have studied the peregrine rogue wave dynamics, with a solitons on finite background (SFB) ansatz, in the recently proposed (Phys. Rev. Lett. **110** (2013) 064105) continuous nonlinear Schrödinger system with parity-time symmetric Kerr nonlinearity. We have found that the continuous nonlinear Schrödinger system with PT-symmetric nonlinearity also admits Peregrine Soliton solution. Motivated by the fact that Peregrine solitons are regarded as prototypical solutions of rogue waves, we have studied Peregrine rogue wave dynamics in the c-PTNLSE model. Upon numerical computation, we observe the appearance of low-intense Kuznetsov-Ma (KM) soliton trains in the absence of transverse shift (unbroken PT-symmetry) and well-localized high-intense Peregrine Rogue waves in the presence of transverse shift (broken PT-symmetry) in a definite parametric regime.

## 1. Introduction

The idea of parity-time (PT)-symmetry in the context of a wide class of non-Hermitian Hamiltonians counter-intuitively having entirely real eigen-spectra once the system obeys parity-time (PT)-symmetry was proposed in the path-breaking work of Bender and Boettcher [1]. Since then, it has witnessed extensive research activities both in theoretical and experimental domains [2-34]. A necessary condition for the Hamiltonian $H = -\frac{1}{2}\frac{d^2}{dx^2} + V(x)$, where we have $V(x) = V_R(x) + i\, V_I(x)$, to be PT-symmetric is that the complex potential $V(x)$ and its real and imaginary parts $V_R(x)$ and $V_I(x)$ respectively obey the following relationships: $V^*(-x) = V(x)$ i.e. $V_R(-x) = V_R(x)$ and $V_I(-x) = -V_I(x)$. In this case, the spectrum of the system is real and the system is said to be in unbroken PT-phase regime, whereas, when the imaginary part of the potential $V_I(x)$ goes beyond a critical value, the spectrum becomes complex or imaginary and the system is said to be in broken PT-phase. Of late, it has been found that optics can provide the platform for investigating PT-symmetry related ideas due to the fact that the paraxial equation of diffraction in optics is isomorphic to the Schrödinger equation in quantum mechanics [2-7] allowing observation of PT-symmetry in optical waveguide structures and lattices [8]. In the numerous studies and investigations, it has been found that the idea of PT-symmetric optics can result in novel optical components and applications [9-19]. On the other front, nonlinear Schrödinger systems with PT-symmetric linear potentials $(i\psi_z + \frac{1}{2}\psi_{xx} + V(x)\psi + |\psi|^2\psi = 0)$ have been studied quite rigorously in the recent past [20-30]. In a nonlinear Schrödinger system, Kerr nonlinearity can dynamically create an effective linear potential which in general may not be PT-symmetric. In consequence, if the even symmetry of the real part of the effective potential gets broken, the system can observe PT-symmetry breaking instability in its wave evolution dynamics. In a recent study, Ablowitz and Musslimani [31] have considered an alternative class of highly nonlocal nonlinear Schrödinger equation where the standard third-order nonlinearity $|\psi|^2\psi$ is replaced by its PT-symmetric form: $\psi(x,z)\psi^*(-x,z)\psi(x,z)$. The study also reveals that this equation is fully integrable since it possesses linear Lax pairs and an infinite number of conserved quantities. In another work, Christodoulides et al. have studied dynamical behaviors of continuous and discrete Schrödinger systems exhibiting parity-time (PT) invariant nonlinearity [32]. The study reveals that the system yields simultaneous bright and dark soliton solutions and that the shift in the transverse co-ordinate 'x' results in the PT-symmetry breaking. This way the wave dynamics of the solitons undergoes instability. It is worthwhile to mention that the idea of parity-time symmetry and the transverse shift, in the context of Rogue waves, have also been explored by a group of researchers [33,34]. Other works in this connection include dark and anti-dark soliton interactions in the nonlocal nonlinear Schrödinger equation with self-induced parity-time-symmetric potential [35], periodic and hyperbolic soliton solutions in nonlocal PT-symmetric equations [36]. On a different side, as a limiting case of a wide class of solutions to the nonlinear Schrödinger equations, Peregrine solitons (PSs) being localized both in evolution and transverse variables draws fundamental importance [37]. It is also regarded as the limiting case of the transverse co-ordinate periodic Akhmediev breather [37-39] or the longitudinal co-ordinate periodic Kuznetsov-Ma (KM) breather [38, 40]. In spite of being theoretically predicted in 1983 by H. Peregrine [41], it was not until 2010 by Kibler et al. [40] that the Peregrine soliton has been experimentally demonstrated in nonlinear fiber optics. After that, Peregrine soliton has been studied in numerous contexts such as: Peregrine solitons in a multi-component plasma with negative ions [42], Peregrine soliton generation and breakup in the standard telecommunications fiber [43], interaction of Peregrine solitons [44], breather-like solitons extracted from the Peregrine rogue wave [45], Peregrine solitons and algebraic soliton pairs in Kerr media [46] and so on. Often, the Peregrine soliton is seen as a rogue-wave prototype [47] for its close resemblance to the rogue-wave dynamics. It is interesting to note that "Rogue waves" or "freak waves" which are large amplitude "waves appearing from nowhere and disappearing without a trace" (WANDT) find its roots in hydro-dynamical systems [48, 49], nonlinear fiber optics [42], Bose-Einstein condensates [50] etc. Optical rogue waves have been studied in singly resonant parametric oscillators [51], mode-locked lasers [52], in optically injected lasers [53]. In this work, we have considered a Peregrine soliton ansatz to the focusing nonlinear Schrödinger equation with anomalous dispersion with parity-time (PT) symmetric nonlinearity. We intend to study the dynamics of the c-PTNLSE under Peregrine soliton or more generally solitons on finite background (SFB) excitation both in unbroken and broken PT-phases. Since the Peregrine solitons have been seen as a prototype of rogue waves, we study the dynamics of initial PS solution in the c-PTNLSE model and see if it can yield Peregrine rogue (PR) wave in the broken PT-phase. This work may encourage the nonlinear physics community to investigate many other nonlinear systems [54-57] starting with quadratic to power law media with such novel type of nonlinearity.

## 2. Theoretical model

In this work, as stated in the previous section, we are considering the nonlocal nonlinear Schrodinger equation, in normalized units, where the standard third order nonlinearity $|\psi(x,z)|^2 \psi(x,z)$ is replaced with its PT symmetric counterpart $\psi(x,z)\psi^*(-x,z)\psi(x,z)$ to obtain:

$$i\psi_z + \tfrac{1}{2}\psi_{xx} + \psi(x,z)\psi^*(-x,z)\psi(x,z) = 0 \quad (1)$$

Here $\psi(x,z)$ is the dimensionless field. $x$ and $z$ are normalized distance and time. It is straightforward to show that Eq. (1) possesses the following solitons on finite background (SFB) solutions:

$$\psi(x,z) = \left[\frac{(1-4a)\cosh(bz)+\sqrt{2a}\cos(\Omega x)+i\,b\,\sinh(bz)}{\sqrt{2a}\cos(\Omega x)-\cosh(bz)}\right] \quad (2)$$

where, $\Omega$ is the dimensionless spatial modulation frequency, $a = 1/2(1 - \Omega^2/4)$ with $0 < a < 1/2$ determines the frequencies experiencing gain and $b = \sqrt{8a(1-2a)}$ is the instability growth parameter. For $a \to 1/2$, the above solution reduces to the rational soliton form i.e. Peregrine soliton:

$$\psi(x,z) = \left[1 - \frac{4(1+2iz)}{1+4x^2+4z^2}\right]e^{iz} \quad (3)$$

## 3. Numerical simulations and analysis

In order to analyze the continuous PTNLSE numerically, in the context of SFB solutions, we have taken the following initial excitation from Eq. (2): $\psi(x,0) = \left[\frac{(1-4a)+\sqrt{2a}\cos(\Omega x)}{\sqrt{2a}\cos(\Omega x)-1}\right]$. Eq. (1) is solved numerically by using the so-called split-operator method. For a detailed discussion of the method readers are referred to Ref. [58]. In a recent work [32] it is shown that introduction of the shift, $\varepsilon$, in the transverse co-ordinate, $x$, may give rise to instability in the wave dynamics. This happens because the self-induced potential $V(x,z)$ starts to have imaginary contribution, due to transverse shift, which once goes beyond a critical value results in instability of the wave dynamics. This motivates us to consider various cases with regard to broken and unbroken PT-phases.

*Without any transverse shift.* We first consider the case without transverse shift. Fig. 1 exhibits the dynamical evolutions of the intensity of the optical field under the SFB excitation. As can be observed from Fig.1 (a)-(b), we find existence of Kuznetsov-Ma (KM) solitons periodic in the evolution variable 'z'. KM soliton is dependent on the modulation parameter $a$. The spatial periodicity of KM solitons keeps on increasing with increase in the modulation parameter '$a$'.

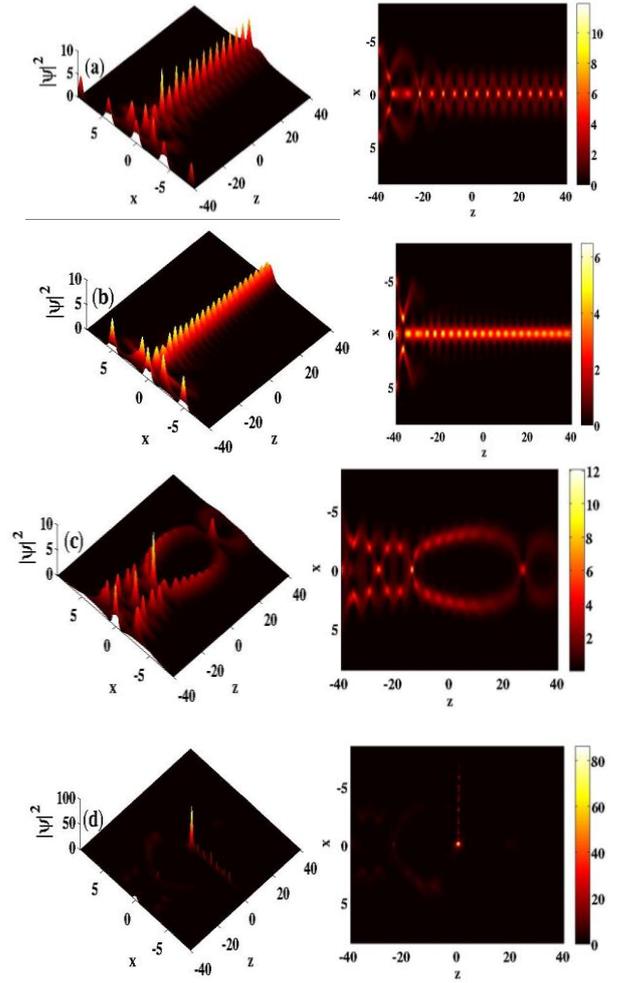

Fig. 1: (Color online) Left hand panel: Evolutions of optical intensity in the z-x plane. Right hand panel: corresponding 3D density plots. (a) $a$=0.20, (b) $a$=0.30, (c) $a$=0.45, (d) $a$=0.47. $\varepsilon = 0.0$.

The spatial periodicity of KM solitons keeps on increasing with increase in the modulation parameter '$a$' and results in the Peregrine solitons as $a \to 1/2$, as could be seen from Fig. 1 (c)-(d). In passing, it is worth noting that the standard NLSE supports Akhmediev breather solutions, which become more and more localized both in time and space co-ordinates with the increase of the modulation parameter '$a$' [40].

*With a small but nonzero transverse shift.* Fig. 2 elucidates the effect of shift in the transverse co-ordinate 'x' on the evolution of the optical field dynamics. From Fig. 2, we can infer that with the increase in the transverse shift parameter, the optical field dynamics tend to be more localized in the z-x plane. The role of the transverse shift is two-fold: first, in triggering the PT-instability, and second, ensuring the nonlinear interactions between localized modes. That is why, in Fig. 2 (a) and (b), although we see presence of a dominant single second-order kind of PS with a wing-like structure, in Fig. 2 (c), we find appearance of another second-order type of PS in the CW background. It is worthwhile to note that a strict mathematical definition of second order PS soliton for the continuous Schrodinger equation considered in this work, unlike some other nonlinear systems [34], is not available. Our assumption of second order PS is based on plots resulting from numerical simulations. More interestingly, in Fig. 2 (d), we finally observe one PS in the CW background. This is simply because in Fig. 2 (d), further increase of the transverse shift acts as a perturbation which gets amplified via modulation instability and through nonlinear wave-mixing processes the multiple PS profiles interact with each other to transfer energy from one to the other resulting into giant Peregrine rogue wave. One crucial point to mention here is that the occurrence of the PRW depends on very definite parameter values, which is quite evident as far as the rarity of the rogue wave phenomenon is concerned. In the corresponding density plot of Fig. 2(d), we also observe occurrence of certain radiation states alongside the Peregrine rogue wave profile.

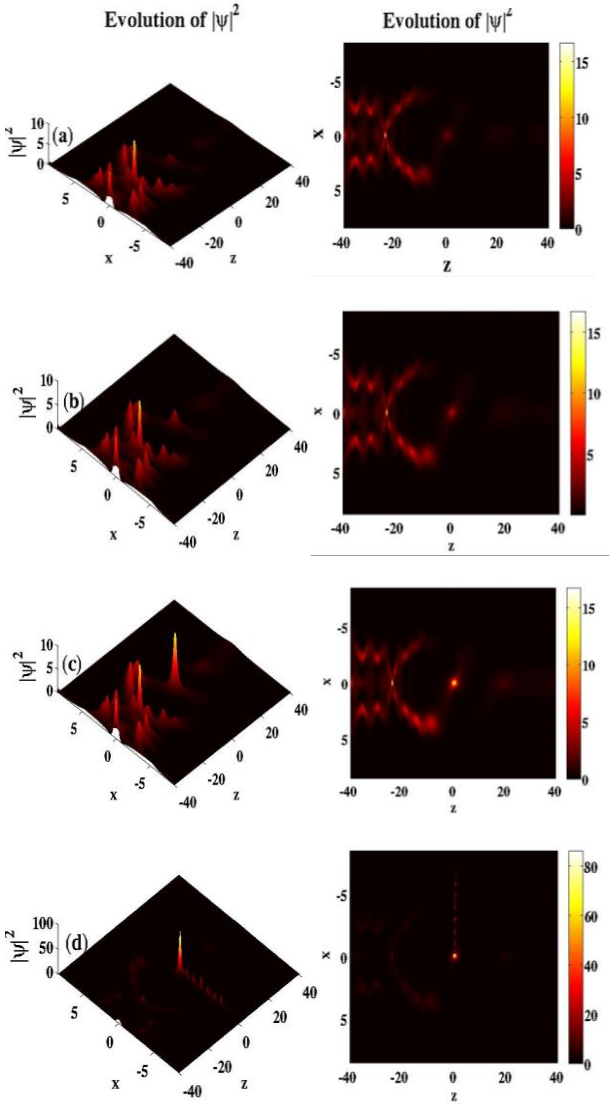

Fig. 2: (Color online) Left hand panel: Evolutions of optical intensity in the z-x plane. Right hand panel: corresponding 3D density plots. (a)$\varepsilon = 2 \times 10^{-6}$, (b)$\varepsilon = 3 \times 10^{-6}$, (c)$\varepsilon = 4 \times 10^{-6}$, (d)$\varepsilon = 4.53 \times 10^{-6}$. Other parameter value: $a = 0.47$.

It is worthwhile to mention here that in a recent work in the context of PT-symmetric coupled waveguides, occurrence of two KM solitons trains hinged upon a second-order Peregrine soliton creating a wing-like shape has been reported [38]. Our study on c-PTNLSE model also reveals existence of similar wing-like KM soliton trains hinged upon a second-order Peregrine soliton as is evident from Fig. 2 (a)-(b).

The fact that inclusion of a nonzero shift in the transverse co-ordinate results in the instability in the wave dynamics, as suggested in [32], could be clearly seen from Fig.3. This instability can be attributed to the induction of an imaginary part in the self-induced potential, which goes beyond a critical value. In Fig. 3 (b) and its corresponding density plot shows the onset of instability in the optical field dynamics in presence of the transverse shift. The intensity of the optical field keeps on increasing with the normalized propagation distance.

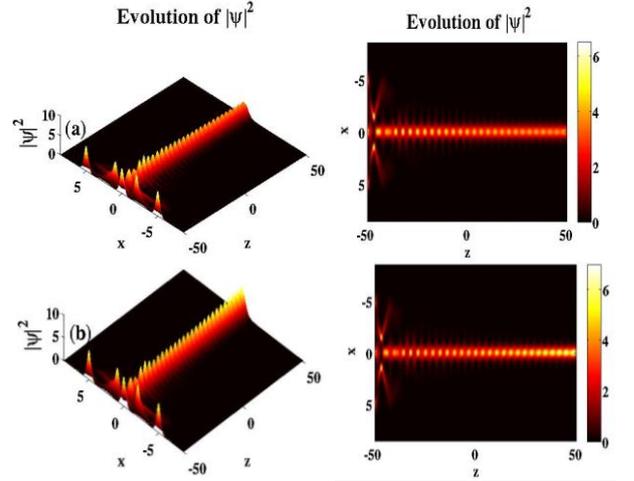

Fig. 3: (Color online) Left hand panel: Evolutions of optical intensity in the z-x plane. Right hand panel: corresponding 3D density plots. (a)$\varepsilon = 0.0$, (b)$\varepsilon = 3 \times 10^{-3}$. Other parameter value: $a = 0.30$

Without any transverse shift, in Fig. 4(a) we observe two lines of peaks of KM soliton train along the z-axis, existing at different locations. These two lines of peaks of KM soliton trains are hinged upon a second order Peregrine soliton. Fig. 4(c) depicts the corresponding power along the propagation distance. Now, quite interestingly, we find that for the same set of parameter values but with a nonzero shift in the transverse co-ordinate, the system supports robust (with respect to the transverse shift parameter in the range $\varepsilon = [4.5, 4.605] \times 10^{-6}$) Peregrine soliton solution with enhanced intensity, strongly localized in the z-x plane, as is clear from Fig. 4 (b) and (d). This enhancement in the intensity of the optical field is owing to the broken PT-symmetry of the self-induced potential. Due to the prototypical analogy between Peregrine soliton and the Peregrine rogue wave, we can qualitatively argue that the enhanced PS dynamics resemble to the Peregrine rogue waves.

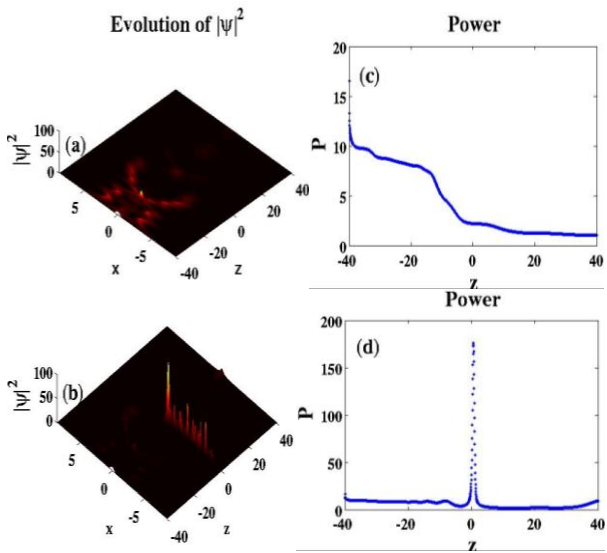

Fig. 4: (Color online) Left hand panel: Evolutions of optical intensity in the z-x plane. Right hand panel: Corresponding power along the propagation distance. (a), (c): $\varepsilon = 0.0$ and (b),(d): $\varepsilon = 4.605 \times 10^{-6}$. Other parameter value: $a = 0.47$.

## 4. Conclusions

In the continuous PTNLSE model, taking the general solitons on the finite background ansatz as the initial excitation, a numerical study has been carried out with special emphasis given to the Peregrine soliton dynamics. Since, the Peregrine solitons have been seen as a rogue-wave prototype for a long time, we numerically confirmed that an initial PS excitation could yield PR wave in the broken PT-phase. Upon numerical computation, we observe the appearance of low-intense Kuznetsov-Ma (KM) soliton trains in the absence of transverse shift and well-localized high-intense Peregrine Rogue waves in the presence of transverse shift in a definite parametric regime. In the earlier case, the PT-symmetry is unbroken, whereas in the later case the PT-symmetry is broken. This work may boost further investigation of optical rogue waves and its dynamics in the context of nonlocal nonlinear systems.